\documentstyle[preprint,aps]{revtex}
\begin{document}
\preprint {WIS-95/22/May-PH}
\draft
\title{The ratio $F_2^n/F_2^p$ from the analysis of data using
a new scaling variable}
\author{S. A. Gurvitz}
\address{Department of Particle Physics, Weizmann Institute of
         Science, Rehovot 76100, Israel}
\maketitle
\begin{abstract}
We analyze the proton and deuteron structure functions at large $x$ using
a recently introduced scaling variable $\bar x$. This variable
includes power corrections to $x$-scaling, and
thus allows us to reach the Bjorken limit at moderate $Q^2$.
Using available data we
extract the ratio $F_2^n(x)/F_2^p(x)$ for $x\leq 0.85$.
Contrary to earlier expectations this ratio tends to the value
$\sim$2/3 for $x\to 1$, which corresponds to the quark model prediction
for equal distributions of valence quarks.
\end{abstract}
\vskip 1cm
\par
According to the quark-parton model, the structure functions of hadrons
in the Bjorken limit ($Q^2={\mbox{\boldmath $q$}}^2 -\nu^2\to \infty$ and
$x=Q^2/2M\nu$=const) are directly related to the parton distributions
q$_i(x)$. For instance
\begin{equation}
F_2(x,Q^2)\to F_2(x)=\sum_i e_i^2x{\rm q}_i(x),
\label{a1}
\end{equation}
where the sum is over the partons, whose charges are $e_i$. In the region
$x\to 1$, the contribution of sea quarks can be neglected. Then assuming the
same distribution of the valence quarks, one easily finds from
Eq. (\ref{a1}) that the neutron-to-proton ratio
$F_2^n(x)/F_2^p(x)$ approaches 2/3 for $x\to 1$. If, however, the quark
distributions are different,
one can establish only upper and lower limits for this ratio,
$1/4<F_2^n/F_2^p<4$, which follow from isospin invariance\cite{nacht1}.
The existing data show considerable $Q^2$-dependence of the structure
functions which is attributed mainly to the QCD logarithmic corrections
to Bjorken scaling.  However, at high-$x$ the scaling
violations are dominated by power corrections $\propto 1/Q^2$
(higher twist and target mass effects), which are difficult to evaluate.
Therefore, in order to check the parton model predictions,
one needs to obtain
the structure functions at high $Q^2$,
where these corrections are small. At present, high $Q^2$ structure
functions ($Q^2\simeq 250$ (GeV/c)$^2$) extracted from
BCDMS\cite{bcd1} and NMC\cite{nmc1} data are available only
for $x\lesssim 0.7$. The ratio  $F_2^n(x)/F_2^p(x)$
obtained from an analysis of these data\cite{bcd2,nmc2} shows steady
decrease with $x$. Thus, it is usually assumed that this ratio would reach
its lower bound, $F_2^n/F_2^p\to 1/4$, for $x\to 1$. This
corresponds to $d(x)/u(x)\to 0$ for $x\to 1$, where $d(x)$ and $u(x)$
are the distribution functions for up and down quarks in the proton.
To check this assumption one needs data for the structure functions
at $x\gtrsim 0.7$. The latter are now
available only at moderate values of momentum
transfer, $Q^2\lesssim 30$ (GeV/c)$^2$\cite{slac1,slac2,slac3,slac4}, and
exhibit very strong $Q^2$-dependence.

A part of the $Q^2$-dependence of the structure functions, generated
by the target mass corrections, is usually
accounted for by using the Nachtmann scaling variable\cite{nacht2,rgp}
\begin{equation}
\xi =\frac{2x}{1+\sqrt{1+4M^2x^2/Q^2}},
\label{aa1}
\end{equation}
instead of the Bjorken variable $x$. The question is whether
the target mass effect is responsible for a major
part of the scaling violation at high $x$. If so, the replacement of
$x$ by $\xi$ in structure functions would allow us to reach the scaling limit
already at moderate $Q^2$. However, the analysis of
recent  high $x$ SLAC data\cite{slac4} in terms of the Nachtmann variable
still reveals strong $Q^2$-dependence of the structure functions.

Besides the target mass effects, one can expect important nonperturbative
effects from the confining interaction of the partons in the final state.
Indeed, the partons are never free, so that
the system possesses a discrete spectrum in the final state.
Although in the Bjorken limit the struck quark can be considered
a free particle, the discreteness of the spectrum manifests itself in
$1/Q^2$ corrections to asymptotic structure functions\cite{green,gr}.
One can expect that these corrections are significant in particular
at high $x$, where lower-lying excitations should play an important role.

We have found\cite{g} that the target mass and confining interaction effects
in the final state can be effectively
accounted for by taken the struck quark to be off-shell,
with the same virtual mass before
and after the virtual photon absorption. As a result, the Bjorken
scaling variable $x$ is replaced by a new scaling variable
$\bar x\equiv\bar x(x,Q^2)$, which is
the light-cone fraction of the {\em off-shell} struck quark.
Explicitly,
\begin{equation}
\bar x=\frac{x+\sqrt{1+4M^2x^2/Q^2}-
\sqrt{(1-x)^2+4m_s^2x^2/Q^2}}
{1+\sqrt{1+ 4M^2x^2/Q^2}},
\label{a2}
\end{equation}
where $M$ is the target mass and $m_s$ is the invariant mass of
spectator partons (quarks and gluons). For $Q^2\to \infty$ or for $x\to 0$
the variable $\bar x$ coincides with the Nachtmann variable $\xi$,
Eq. (\ref{aa1}). However, at finite
$Q^2$ these variables are quite different.

It follows from Eq. (\ref{a2}) that $\bar x$ depends on the invariant
spectator mass, $m_s$. The latter can be considered a function
of the external parameters only\cite{g}. In the limit $x\to 1$
(elastic scattering) no gluons are emitted, and thus $m_s\to m_0$,
the mass of a two-quark system (diquark). When $x<1$, the spectator mass
$m_s$ increases due to gluon emission.
For $x$ close to 1, $m_s$ can be
approximated as
\begin{equation}
m_s^2\simeq m_0^2+C(1-x),
\label{a3}
\end{equation}
where the coefficient $C\sim$ (GeV)$^2$. In the following we regard
it as a phenomenological parameter, determined from the data.

Consider the proton and deuteron structure functions (per nucleon)
$F_2^{p,d}(x,Q^2)$ at large $x$. These are shown
in Fig. 1a,b as functions of $\bar x$ for $Q^2$=230 (GeV/c)$^2$, which
is the maximal value of $Q^2$ in the BCDMS measurements\cite{bcd1}.
(Notice that for such high values of $Q^2$ the variable
$\bar x$ is close to $x$, Eq. (\ref{a2})). The data are taken
from\cite{bcd1}, where
the solid lines correspond to a
15 parameter fit to BCDMS and NMC data\cite{nmc1}.
For smaller values of $Q^2$,
the violations of Bjorken scaling are very significant, Fig. 1c,d.
By assuming that the major part of $x$-scaling violations at high $x$
are correctly accounted for by the variable $\bar x$,
the $Q^2$-dependence of the structure functions in this region
is given by
\begin{equation}
F_2^{p,d}(x,Q^2)=F_2^{p,d}(\bar x(x,Q^2)).
\label{a4}
\end{equation}
Let us compare this
result with two BCDMS and SLAC data bins for $x$=0.65, 0.75\cite{bcd1,slac1},
Fig. 1c,d, which are the largest values of $x$ available in the BCDMS
experiment. We find that these data are perfectly
reproduced (the dashed lines in Fig. 1c,d),
by taking the spectator mass  $m_s^2$=0.75 (GeV)$^2$ for
$x$=0.75 and $m_s^2$=1.05 (GeV)$^2$ for $x$=0.65. It is quite remarkable that
the same values of $m_s$ are obtained for
proton and deuteron targets, although the corresponding structure functions
are rather different. Using Eq. (\ref{a3}) one finds that these values of
$m_s$ determine the parameters $m_0$ and $C$, namely $C=3$ (GeV)$^2$
and $m_0=0$. The latter implies
that the spectator quarks are massless and collinear.

Using these values of $m_0$ and $C$ for definition of $m_s$ in Eq. (\ref{a2})
we can study the structure functions for $x>0.75$.
Consider first the data for the proton structure function from the
SLAC experiments\cite{slac1,slac2,slac3,slac4} in the region $x > 0.6$ for
$5\lesssim Q^2\lesssim30$ (GeV/c)$^2$,
plotted as a function of $\xi$ and $\bar x$
respectively, Fig. 2a,b. The data points close
to the region of resonances were excluded by a requirement on the
invariant mass of the final state, namely
$(M+\nu )^2-{\mbox{\boldmath $q$}}^2>
(M+\Delta)^2$, where $\Delta=$300 MeV.
The solid line and the three data points
in Fig. 2a,b correspond to the asymptotic structure
function at $Q^2$= 230 (GeV/c)$^2$, the same as in Fig. 1a.
The analysis in terms of the Nachtmann scaling variable $\xi$, Fig. 2a,
shows poor scaling. Moreover, the data points are far off
the asymptotic structure function (the solid line).
In contrast, the same data plotted as a function of $\bar x$,
Fig. 2b, show excellent scaling. Also the data points
completely coincide with the
structure function at $Q^2$= 230 (GeV/c)$^2$, available for $x<0.75$.
This agreement provides strong evidence
that the $\bar x$-scaling is not accidental. We therefore propose
that the data points in Fig. 2b represent a measurement of the
asymptotic structure function for $x>0.75$ as well.

Next, consider the deuteron structure function from the SLAC
data\cite {slac1,slac2,slac3}, Fig. 3a,b.
As in the previous case, we exclude the region of resonances by
taking the invariant mass in the final
state greater then $M+\Delta$, where
$\Delta$=300 MeV. In addition, in order to avoid complications from
binding and Fermi motion effects, we exclude from our analysis
the data points with $x>0.9$.
(This restriction is relevant only for the data\cite{slac3}).
Indeed, recent calculations of Melnitchouk
{\em et al.}\cite {mel} show that
the ratio $2F_2^d/(F_2^p+F_2^n)$ is about 1.13 for
$x=0.9$ and $Q^2$=5 (GeV/c)$^2$, and it rapidly increases for $x>0.9$.
However, for $x<0.85$, this ratio is within
5\% of unity\cite{not1}.
Fig. 3a shows the deuteron data plotted as a function of the Nachtmann
variable $\xi$.
The solid line and three data points show the structure function at
$Q^2$= 230 (GeV/c)$^2$,
the same as in Fig. 1b. One finds that the data display no scaling
and they are far from the
solid line. The same data as a function of $\bar x$ are shown in Fig. 3b.
As in the proton case the data show very good scaling
and do coincide with the structure function
at $Q^2$= 230 (GeV/c)$^2$.
Unfortunately, there are no deuteron data in the high-$x$
region for large $Q^2$, as for instance the proton data\cite{slac4}.
As a result, the available deuteron data
allow us to determine the asymptotic structure function only up to $x=0.85$.

Now with the asymptotic structure functions $F_2(\bar x)=
F_2(x)\equiv F_2(x,Q^2\to\infty)$ found above, we can
obtain the ratio $F_2^n(x)/F_2^p(x)$.  For this purpose we parametrize
the asymptotic structure functions as
$F_2^{p,d}(x)=\exp (-\sum_{i=0}^4 a_ix^i)$
and determine the parameters $a_i$ from the best fit to the data in
Figs. 2b, 3b. The resulting $F_2^n/F_2^p$ ratio is shown
in Fig. 4 by the solid line. The dotted lines are the error bars
on the fit, which combine
statistical and systematic uncertainties.
The dashed line corresponds to $F_2^n/F_2^p=2/3$.
For a comparison, we show by the dot-dashed line
a polynomial extrapolation of this ratio to large $x$,
obtained from BCDMS and NMC data by assuming that $F_2^n/F_2^p\to 1/4$
for $x\to 1$\cite{bcd2}.

Our results shown in Fig. 4 demonstrate that contrary to earlier
expectations, the ratio
$F_2^n/F_2^p$ does not approach its lower bound, but increases up to
approximately 2/3. The latter is the
quark model prediction, assuming identical distributions for each
of the valence quarks.
The accuracy of our results will be checked in future experiments,
which will provide high $Q^2$ data for the structure functions at
large $x$.

\section{Acknowledgments}

I am grateful to A. Bodek and S. Rock for providing me with
data files for proton and deuteron structure functions. Special
thanks to B. Svetitsky for reading the manuscript and making valuable
comments on it.

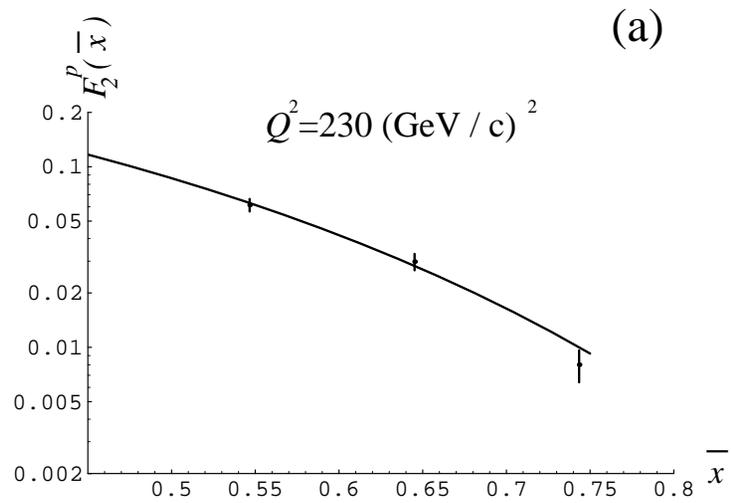
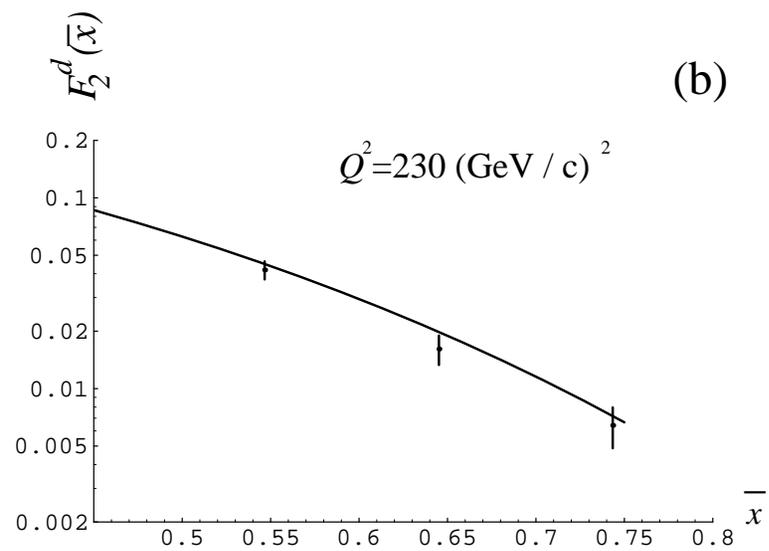
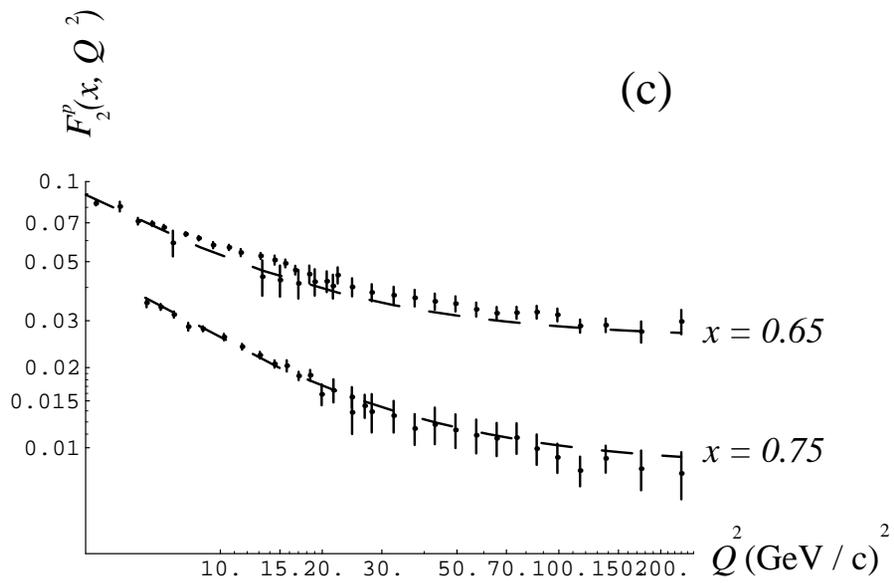
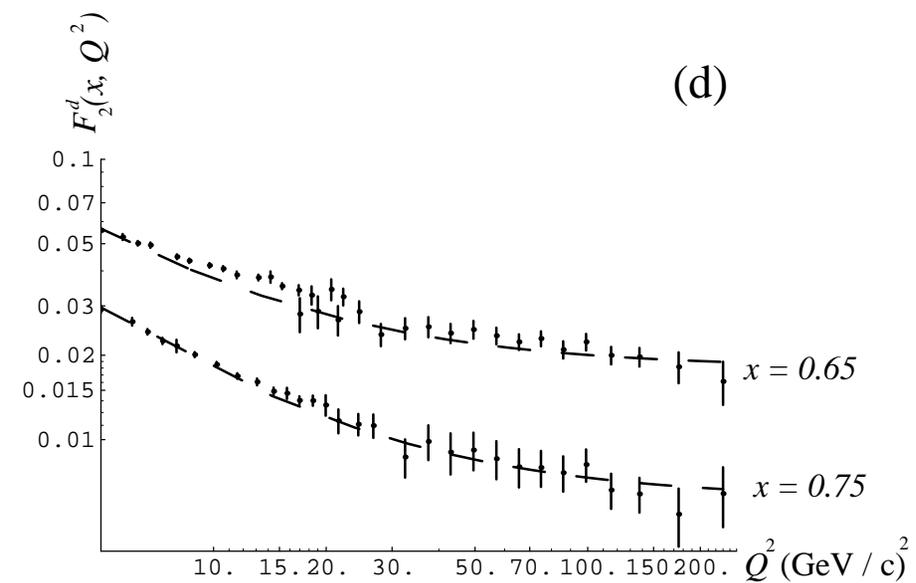
\begin{figure}[tb]
\caption[]{(a,b)$\;F_2^{p,d}(\bar x)=
F_2^{p,d}(\bar x, Q^2)$ at $Q^2$=230 (GeV/c)$^2$.
The data are from BCDMS measurements\cite{bcd1}.
The solid line is the 15 parameter
fit\cite{nmc1}  to BCDMS, NMC and SLAC data.
(c,d) Proton and deuteron structure functions
at constant $x$. The dashed lines show $Q^2$-dependence of the
structure functions given by Eq. (\ref{a4}).
The data are from BCDMS\cite{bcd1} and SLAC\cite{slac1,slac2} experiments.
The error bars show combined statistical and systematic errors.
\label{fig1}}
\end{figure}

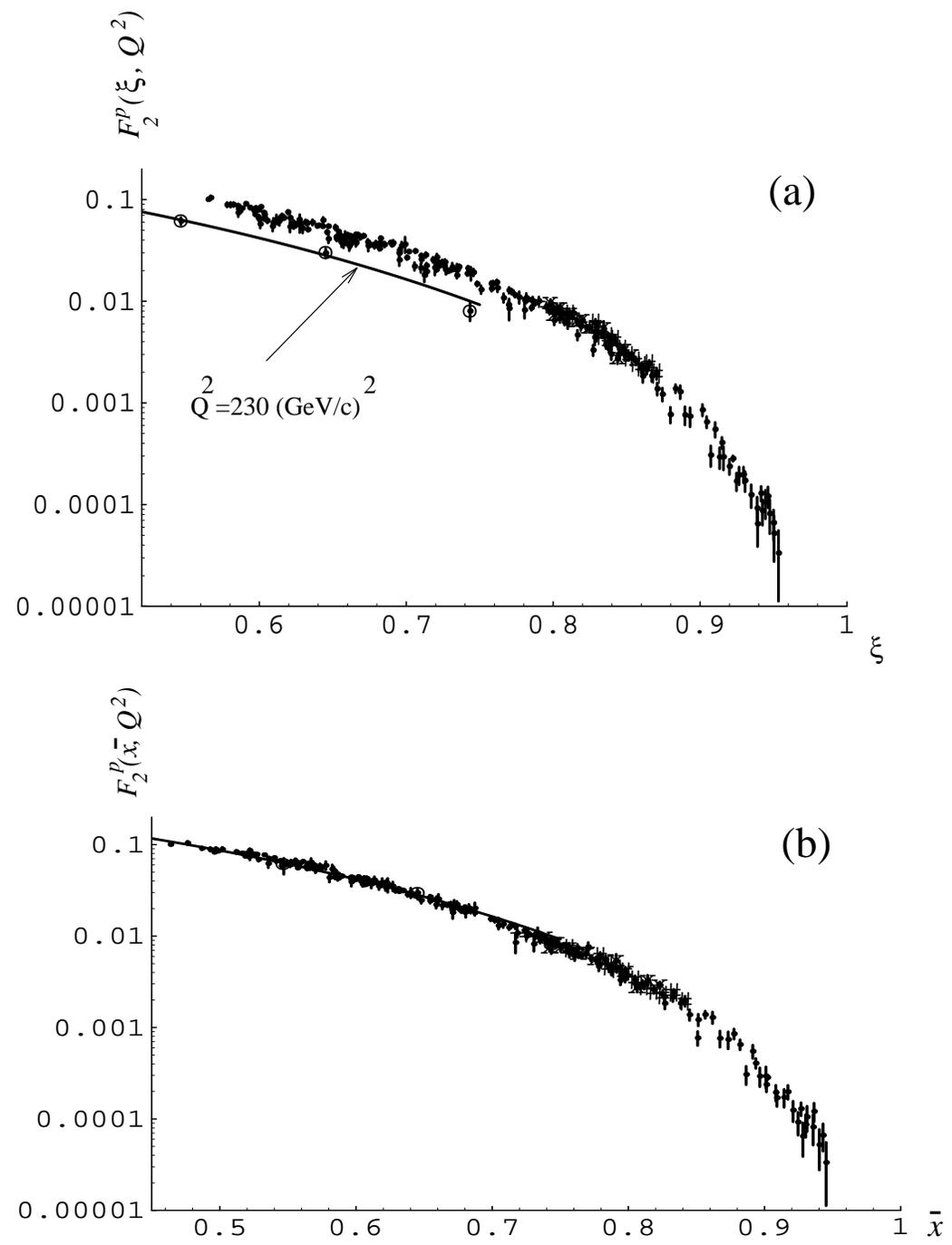
\begin{figure}[tb]
\caption[]{(a) SLAC data for the proton structure function for
$5\lesssim Q^2\lesssim 30$ (GeV/c)$^2$,
plotted as a function of the Nachtmann variable $\xi$, Eq. (\ref{aa1}).
The data with largest value of $\xi$ are taken from recent
measurements\cite{slac4}. Three high-statistics data sets
for $Q^2$=5.9, 7.9, and 9.8 (GeV/c)$^2$, taken from\cite{slac3,slac4}, are
marked by ``+", ``x", and ``{\#}" respectively. The other data points
are from\cite{slac1,slac2}. The error bars show combined statistical
and systematic uncertainties. Three data points marked by ``o" and the
solid curve are the same as in Fig. 1a, and
show the structure function at $Q^2$=230 (GeV/c)$^2$.
(b). The same data plotted as a function
of the scaling variable $\bar x$. The data points for
$5\lesssim Q^2\lesssim 30$ (GeV/c)$^2$ coincide well with
the structure function at $Q^2$=230 (GeV/c)$^2$.
\label{fig2}}
\end{figure}

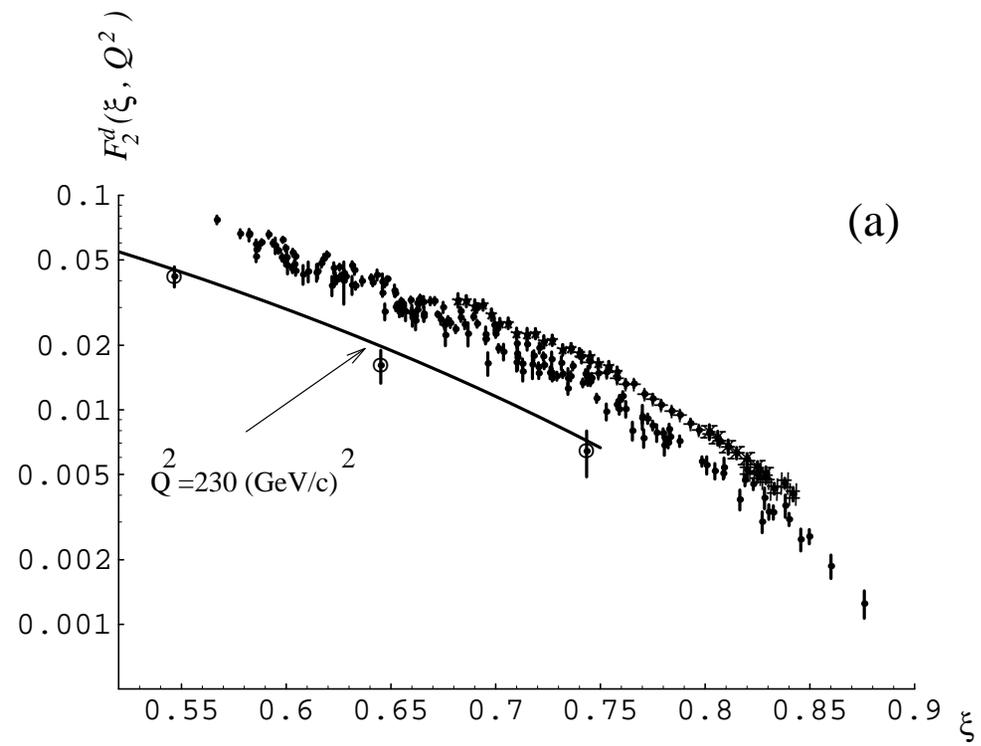
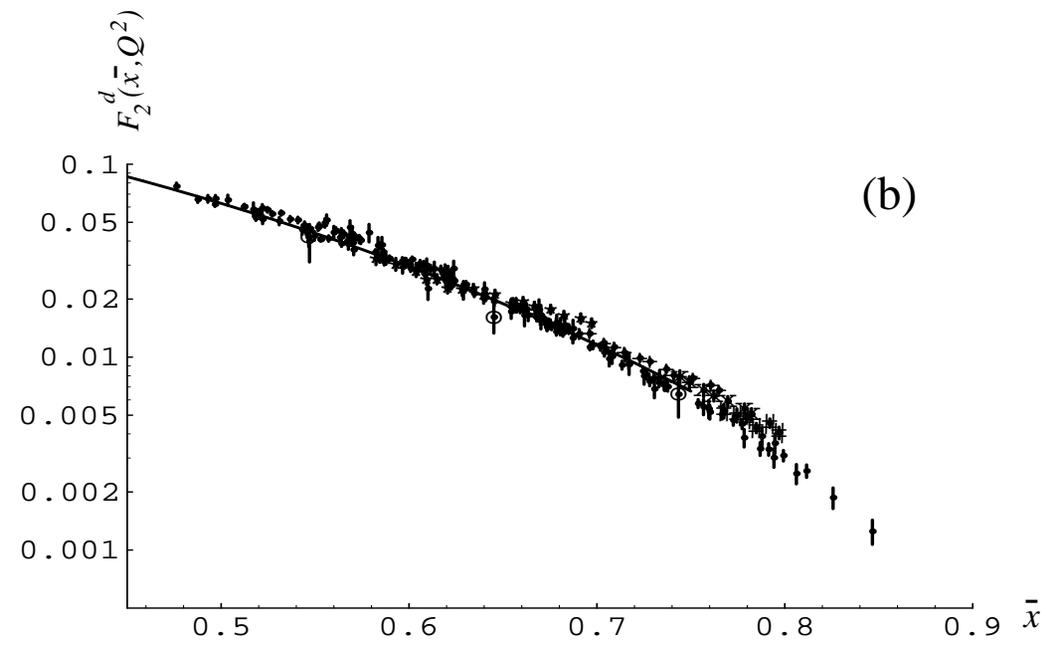
\begin{figure}[tb]
\caption[]{(a) SLAC data for deuteron structure function for
$5\lesssim Q^2\lesssim 30$ (GeV/c)$^2$
plotted as a function of the Nachtmann variable $\xi$, Eq. (\ref{aa1}).
Four high-statistics data sets
for $Q^2$=3.9, 5.9, 7.9, and 9.8 (GeV/c)$^2$, taken from\cite{slac3}, are
marked by ``*", ``+", ``x" and ``{\#}" respectively. The other data points
are from\cite{slac1,slac2}.  The error bars show combined statistical
and systematic uncertainties. Three data points marked by ``o" and the
solid curve are the same as in Fig. 1a, and
show the structure function at $Q^2$=230 (GeV/c)$^2$.
(b). The same data plotted as a function
of the scaling variable $\bar x$.
\label{fig3}}
\end{figure}

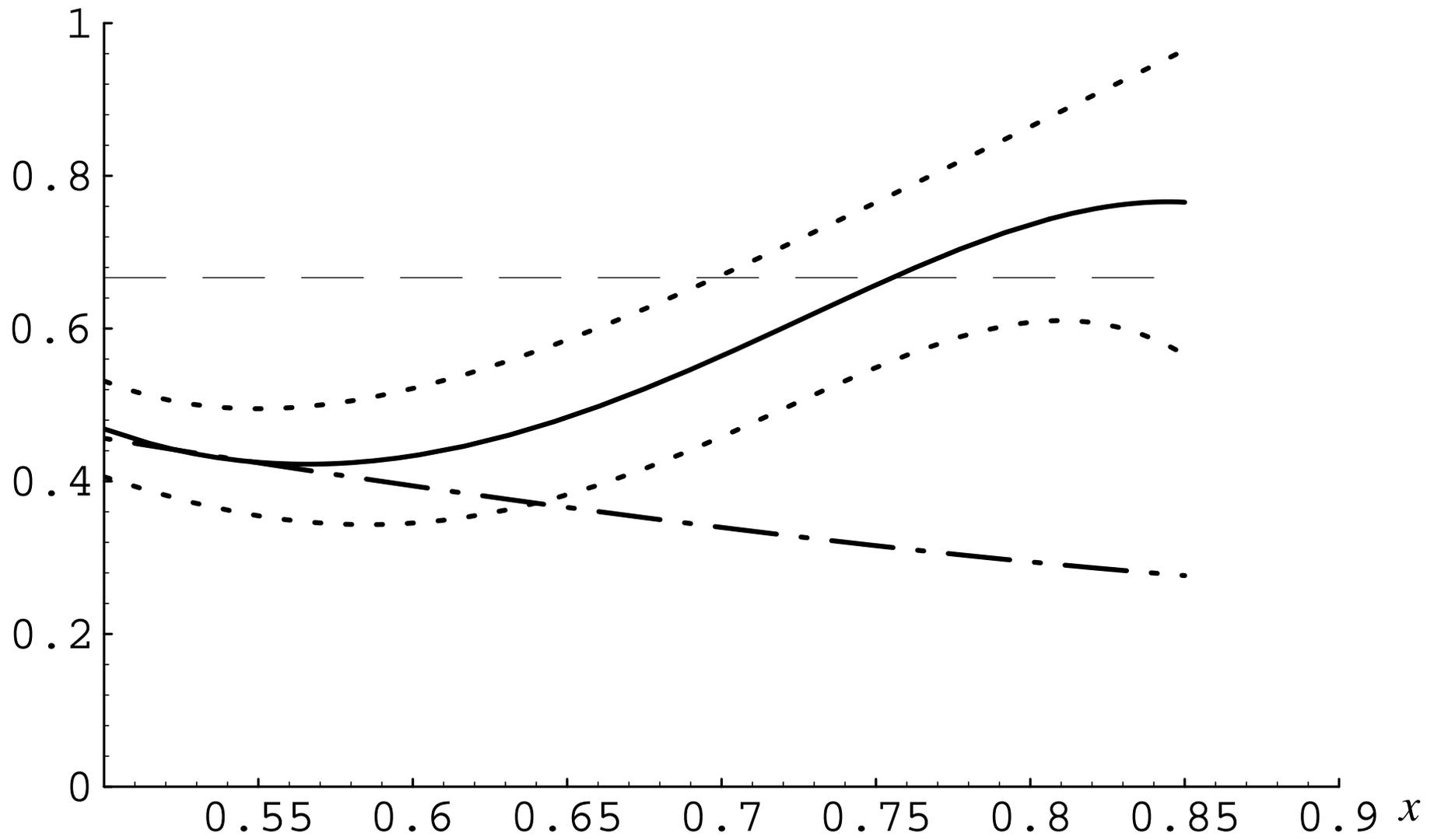
\begin{figure}[tb]
\caption[]{Neutron-to-proton structure function ratio at large $x$.
The solid line is the result of our analysis.
The dotted lines show combined statistical and systematic errors.
The dashed line is the quark model prediction for $x\to 1$ for equal
distributions of valence quarks. The
dot-dashed line shows the expected behavior of this ratio from polynomial
extrapolation of BCDMS and NMC data\cite{bcd2}.
\label{fig4}}
\end{figure}
\end{document}